\documentclass{aa}  

\usepackage{graphicx}
\usepackage{amsmath}
\usepackage{float} 
\usepackage{natbib}
\usepackage{txfonts}
\newcommand\aov{\ifmmode{\alpha_{\rm ov}}\else $\alpha_{\rm ov}$\fi}
\newcommand\amix{\ifmmode{\alpha_{\rm MLT}}\else $\alpha_{\rm MLT}$\fi}

\begin{document} 

\title {The dependence of convective core overshooting on stellar mass}


\author{
A.\ Claret\inst{\ref{inst1}}
\and
G.\ Torres\inst{\ref{inst2}}
}

\offprints{A.\ Claret, e-mail: claret@iaa.es}

\institute{
Instituto de Astrof\'{\i}sica de Andaluc\'{\i}a, CSIC, Apartado 3004, 18080 Granada, Spain\label{inst1}
\and
Harvard-Smithsonian Center for Astrophysics, 60 Garden St., Cambridge, MA 02138, USA\label{inst2}
}

\date{Received; accepted; }

\abstract
{ Convective core overshooting extends the main-sequence lifetime of a
  star. Evolutionary tracks computed with overshooting are quite
  different from those that use the classical Schwarzschild criterion,
  which leads to rather different predictions for the stellar
  properties. Attempts over the last two decades to calibrate the
  degree of overshooting with stellar mass using detached double-lined
  eclipsing binaries have been largely inconclusive, mainly due to a
  lack of suitable observational data. }
{ Here we revisit the question of a possible mass dependence of
  overshooting with a more complete sample of binaries, and examine
  any additional relation there might be with evolutionary state or
  metal abundance $Z$. }
{ We use a carefully selected sample of 33 double-lined eclipsing
  binaries strategically positioned in the H-R diagram, with accurate
  absolute dimensions and component masses ranging from 1.2 to
  4.4~$M_{\sun }$.  We compare their measured properties with stellar
  evolution calculations to infer semi-empirical values of the
  overshooting parameter \aov\ for each star. Our models use the
  common prescription for the overshoot distance $d_{\rm ov} = \aov
  H_p$, where $H_p$ is the pressure scale height at the edge of the
  convective core as given by the Schwarzschild criterion, and \aov\ is
  a free parameter. }
{ We find a relation between \aov\ and mass that is defined much more
  clearly than in previous work, and indicates a significant rise up
  to about $2~M_{\sun}$ followed by little or no change beyond this
  mass. No appreciable dependence is seen with evolutionary state at a
  given mass, or with metallicity at a given mass despite the fact
  that the stars in our sample span a range of a factor of ten in
  [Fe/H], from $-1.01$ to $+0.01$. }
   {~ }

\keywords{ stars: evolution; 
stars: interiors;  
stars: eclipsing binaries }

\titlerunning{Core overshooting}

\maketitle

%

\section{Introduction}
\label{sec:introduction}

Convective core overshooting refers to an extension of the stellar
core beyond the boundaries defined by the classical Schwarzschild
criterion. This criterion for stability against convection requires
the acceleration of the convective elements to vanish, but their
velocities will not necessarily vanish because of inertia. As a
result, there will be penetration of convective cells into the stable
layers above the core.  This extra mixing leads to stellar models with
longer main-sequence lifetimes, as more fuel is available within the
core.  Additionally such models have a higher degree of mass
concentration toward the center, which has observable effects as it
can influence the rate of apsidal motion in close eccentric binary
systems.  Other mechanisms may increase the size of the stellar core
as well, such as internal gravity waves or rotation. For the purposes
of this work we will refer to the increase in the convective core size
simply as core overshooting.

Pioneering theoretical studies on the subject of overshooting date
back more than five decades \citep[see, e.g.][]{Roxburgh:65,
  Saslaw:65}. It has become common to characterize the distance
$d_{\rm ov}$ to which convective elements penetrate beyond the
classical Schwarzschild core by defining an overshooting parameter
\aov\ such that $d_{\rm ov} = \aov\ H_p$, in which $H_p$ is the
pressure scale height. The following list gives examples of the
overshooting parameters adopted in various published grids of stellar
evolution models:

\begin{itemize}   

\item \cite{Schaller:92}: $\aov = 0.20$;

%

\item \cite{Demarque:04}: overshooting starts at $1.2~M_{\sun}$ and
  ramps up to a maximum of $\aov = 0.2$ at $1.4~M_{\sun}$ and larger
  masses (metallicity dependent);

\item \cite{Claret:04}: $\aov = 0.20$;

\item \cite{Pietrinferni:04}: $\aov = (M - 0.9~M_{\sun})/4$ for masses
  $M$ in the range 1.1--1.7~$M_{\odot}$, and $\aov = 0.2$ beyond
  $1.7~M_{\odot}$;

\item \cite{Mowlavi:12}: $\aov = 0.05$ between 1.25--1.70~$M_{\sun}$,
  and $\aov = 0.10$ for $M > 1.7~M_{\sun}$;

\item \cite{Bressan:12}: overshooting parametrized in terms of a scale
  parameter $\Lambda_{\rm c}$, which ramps up linearly between 1.1 and
  $1.4~M_{\sun}$ (metallicity dependent) to a value of 0.5,
  corresponding approximately to $\aov = 0.25$, and remains constant
  for larger masses.

\end{itemize}

Other authors have used alternate formulations that take into account
the influence of the radiation pressure on the extension of the
convective core \citep{Pols:95}, prescriptions with a different
adjustable parameter involving an exponential function that depends on
the size of the classical core and on the pressure scale height
\citep{Paxton:11}, or implementations of the Roxburgh criterion
\citep{Roxburgh:78, Roxburgh:89} to model the extent of the convective
core, also using a different free parameter \citep{VandenBerg:06}.

An important question one may naturally ask regarding overshooting is
whether the extension of the convective core depends on stellar mass,
as has often been assumed (but not always; see above), and if so, how.
This is the main subject of this paper. An even more basic question,
actively debated beginning 25 years ago, is whether overshooting is
needed at all in order to fit the observations. On this second issue
opinions were initially divided: \cite{Andersen:90} provided strong
evidence based on moderately evolved detached eclipsing binary stars
of intermediate mass and several open clusters that some degree of
overshooting is required. On the other hand, \cite{Stothers:91,
  Stothers:92} suggested that observations are adequately fitted with
little or no need for core overshooting.  Subsequent investigations
again supported the need for extra mixing \citep[][see references
  therein]{Claret:91, Schaller:92, Bressan:92}, and further studies of
double-lined eclipsing binaries (DLEBs) presented additional evidence
in the same direction by comparing stellar models with accurately
measured absolute dimensions of the components
\citep[e.g.,][]{Ribas:99, Lastennet:02}.  A similar conclusion was
reached by \cite{Claret:10}, also based on DLEBs as well as the
apsidal motion test.  Recent studies of open clusters have continued
to support the need for extra mixing \citep[see, e.g.,][]{Mowlavi:12},
as have numerous studies of individual eclipsing binaries. Virtually
all modern series of stellar evolution calculations now include some
degree of overshooting, although vestiges of earlier hesitations are
perhaps reflected in that several of these grids still offer
``standard'' models with no overshooting (at least for solar
composition), often used for comparison purposes.

Investigating the possible dependence of overshooting on stellar mass
by means of binaries, as we set out to do here, requires not only an
accurate knowledge of the component masses, but also of the stellar
radii ($R$) and the effective temperatures ($T_{\rm eff}$) for a
meaningful comparison with stellar evolution models. These properties
are typically best determined in detached DLEBs. The sample of such
binaries with the most accurate determinations of their absolute
dimensions has increased steadily in size in the last few decades,
from 45 systems compiled by \cite{Andersen:91} with mass and radius
uncertainties better than about 3\%, to more than twice that number in
the more recent review by \cite{Torres:10}.  A study of the
correlation between \aov\ and mass by \cite{Ribas:00} used a total of
eight DLEBs with component masses in the range 2--12~$M_{\sun}$, and
found a strong dependence of the extra mixing on mass. They relied
very heavily on the massive binary V380~Cyg ($M_1 \sim 11~M_{\sun}$,
$M_2 \sim 7~M_{\sun}$) to establish the slope of the
relation. However, fitting the measured properties of this critical
system has always been problematic, as discussed at length by
\cite{Claret:07} and more recently also by \cite{Tkachenko:14}. The
latter authors reported a new set of precise determinations of the
mass, radius, temperature, chemical composition, and other properties
of the components that have nevertheless remained difficult to
reconcile with models.

The study of \cite{Claret:07} revisited the mass dependence of
\aov\ on the basis of masses, radii, and temperatures of thirteen
DLEBs between about 1.3 and $27~M_{\sun}$, but chose to use the ratio
of the component effective temperatures rather than the individual
temperatures themselves, arguing that the ratios can be determined
more accurately from the light curve analyses. Models were computed
for the exact masses measured in each case. The main conclusion of
that investigation was that the dependence of overshooting on mass is
more uncertain and less pronounced than that proposed by
\cite{Ribas:00}.

More recently there has been renewed interest in this subject,
although the results of new studies have been somewhat inconsistent.
\cite{Meng:14} investigated four DLEBs with component masses of
1.4--3.5~$M_{\sun}$, and found no significant dependence of
overshooting with mass.  \cite{Stancliffe:15} modeled twelve DLEBs
between 1.3 and $6.2~M_{\sun}$, and found that the nine for which
their models provided satisfactory fits to the observations also
showed no evidence for a trend of the extent of overshooting with
mass. \cite{Valle:16} examined simulated DLEBs over a more limited
mass range (1.1--1.6~$M_{\sun}$) and only for evolutionary stages up
to central hydrogen depletion. Rather than addressing the mass
dependence issue directly, they took a step back from the empirical
studies of others and focused instead on quantifying the uncertainties
in deriving \aov\ that depend on the observational uncertainties, the
adopted helium-to-metal enrichment ratio $\Delta Y/\Delta Z$, and
other details of the models such as the amount of diffusion and the
mixing length parameter \amix. They cautioned that some of these can
lead to significant biases in the inferred efficiency of convective
overshooting from stellar models.  \cite{Deheuvels:16} dispensed with
binaries altogether and used diagnostics from asteroseismology of
single stars observed by the {\it Kepler} spacecraft to estimate the
extent of the extra mixing for eight relatively low-mass stars
(1.3--1.5~$M_{\sun}$) in which this was possible. They reported hints
that \aov\ may depend on mass even over this narrow range. Similar
seismic studies of more massive stars \citep[$M >
  7~M_{\sun}$;][]{Aerts:13, Aerts:15} reported no obvious mass
dependence.

In the present paper we again invoke DLEBs to investigate the
dependence of overshooting on stellar mass along the lines of the
\cite{Claret:07} study, though with some important differences:
\emph{i)} We employ a significantly enlarged sample (33 binaries
rather than 13) with a range of stellar masses that spans the entire
regime over which current stellar evolution models ramp up the
efficiency of overshooting from zero, and extends to masses well
beyond those at which models typically consider \aov\ to no longer
increase; \emph{ii)} we apply a more careful selection to include only
systems that are evolved enough for the effects of overshooting to be
discernible; and \emph{iii)} we focus on binaries with the best
measured masses, radii, and effective temperatures, as well as
chemical compositions, when available.

The paper is organized as follows. In Section~\ref{sec:sample} we
state our selection criteria and present our observational sample. The
stellar evolution models (Granada series) and methodology to infer
values of \aov\ for each star in each binary are described in
Section~\ref{sec:methods}. The main results concerning the mass
dependence of \aov\ are given in Section~\ref{sec:results}, followed
by a discussion of the significance of our findings in
Section~\ref{sec:discussion}. Concluding remarks may be found in
Section~\ref{sec:conclusions}. Finally, in Appendix~A we apply the
virial theorem in the framework of extra mixing to investigate the
size of the enlarged convective cores.

\section{Observational sample}
\label{sec:sample}

A key requirement for our study of overshooting as a function of
stellar mass is that the binaries in our sample must have precise
\emph{and} accurate masses, radii, and effective temperatures, and
ideally abundance analyses as well. The most recent critical
compilation of absolute dimensions for normal stars by
\cite{Torres:10} included 95 systems with mass and radius
uncertainties under 3\%, but not all are suitable for our
purposes. This is because a second important requirement is that the
stars must be sufficiently evolved to phases where overshooting has a
large enough influence so that it can be estimated reliably by fitting
models (late stages of the main sequence, or giant phases). Unevolved
binaries near the zero-age main sequence (ZAMS) carry no useful
information on \aov. Fewer than a dozen systems from \cite{Torres:10}
meet this condition, and almost all are still on the main sequence. In
the interim a number of other well-studied detached eclipsing binary
systems have been reported, most notably several located in the Large
and Small Magellanic Clouds (LMC, SMC) containing giant stars. These
objects were discovered in the course of the Optical Gravitational
Microlensing Experiment
(OGLE)\footnote{http://ogle.astrouw.edu.pl/~.}, and have served to
establish a precise distance scale to those galaxies
\citep[e.g.,][]{Pietrzynski:10, Pietrzynski:13, Graczyk:12,
  Graczyk:14}. They are also ideal for our purposes because they are
highly evolved. Although some of them exceed our desired 3\% tolerance
in the radius errors, we have chosen to include those slightly less
precise results because of their high value for this investigation and
because they are often accompanied by a metallicity determination,
which is relatively rare for eclipsing binaries.

\addtocounter{table}{1}

The 33 systems we have selected are listed in Table~\ref{tab:sample},
sorted by decreasing primary mass. One ($\alpha$~Aur) is an
astrometric-spectroscopic binary rather than an eclipsing one, but it
still has useful radius determinations obtained from angular diameter
measurements \citep[see][]{Torres:15} along with accurate temperatures
and a detailed abundance analysis. As described later, in some of our
binaries we have swapped the primary and secondary identifications
relative to those published, after verifying that only this assignment
yields reasonable model fits with the ``primary'' being more evolved
than the ``secondary''. These systems are flagged in the table. A
handful of our objects (most notably YZ~Cas, HY~Vir, $\chi^2$~Hya, and
VV~Crv) are especially valuable in that they have mass ratios very
different from unity, providing greater leverage for fitting models.
Three others happen to have primary components that are pulsating
stars (classical Cepheids).  Finally, it is worth keeping in mind that
temperatures and metallicities are less fundamental than masses and
radii, they are typically more difficult to determine or depend on
external calibrations, and they may be subject to systematic errors in
excess of the formal uncertainties listed in Table~\ref{tab:sample}.

\section{Stellar models and methodology}
\label{sec:methods}

For this work we used the Granada stellar evolution code of
\cite{Claret:04, Claret:12} to generate evolutionary tracks for the
measured masses of the components of our binaries, which span the
range 1.2--4.4~$M_{\sun}$. We fitted these models to the observations
for each system (radii, temperatures, and metallicity when available)
in order to infer \aov\ separately for each star under a number of
assumptions described below.  We note that the abundance-related
quantity most often measured and reported for stars is [Fe/H], whereas
the models are usually parametrized in terms of the overall
metallicity $Z$. Even if [Fe/H] is accurate, the conversion to $Z$
usually assumes that abundances of other elements scale in the same
way as in the Sun, which is not necessarily true in all cases.
Additionally, a complete specification of the composition of the
models requires knowledge of the hydrogen ($X$) or helium ($Y$)
content as well, or equivalently, the adoption of an enrichment law
relating $Y$ and $Z$.  Here we have adopted a primordial helium
content of $Y_p = 0.24$, and a slope for the enrichment law given by
$\Delta Y/\Delta Z = 2.0$. Convective core overshooting in our models
is implemented as described in Section~\ref{sec:introduction},
expressing the extra distance traveled by convective elements beyond
the limits of the core as $d_{\rm ov} = \aov\ H_p$, with $H_p$ being
the pressure scale height at the edge of the convective core as given
by the Schwarzschild criterion. To compute the size of the mixed core,
$r_{\rm ov}$, and to avoid dealing with a region of extra mixing
larger than the classical core, we adopt the following simple
(step-function) algorithm: if $r_c$ (classical radius) is smaller than
$H_p$, the size of the new core is given by $r_{\rm ov} = r_c +
\aov\ H_p$; otherwise, $r_{\rm ov} = r_c (1 + \aov)$.  The region
beyond the classical core is fully mixed and the corresponding
temperature gradient is assumed to be adiabatic.

For stars with convective envelopes we employ the standard
mixing-length formalism \citep{Vitense:58}, \amix\ being a free
parameter. Rotation has not been considered for this work.
High-temperature opacities were taken from the tables provided by
\cite{Iglesias:96}; for lower temperatures we used the tables of
\cite{Ferguson:05}. The element mixture adopted in our models is
essentially that of \cite{Grevesse:98}, giving a solar metallicity of
$Z_{\sun} = 0.0189$. Mass loss was accounted for with the prescription
of \cite{Nieuwenhuijzen:90} for all models except those for red giants
with masses smaller than $4~M_{\sun}$; for the latter we adopted the
formalism of \cite{Reimers:77}. Additional details about the code used
to generate our models can be found in the work of \cite{Claret:04,
  Claret:12}.

For each system in our sample we computed a large grid of evolutionary
models for the measured masses, with overshooting parameters \aov\ for
each component covering the range 0.00--0.40 in steps of 0.05, as well
as mixing length parameters \amix\ between 1.0 and 2.0, in steps of
0.1 (99 models in all, for each binary component). Each track
contained several thousand time steps, and the calculations were done
for the observed chemical composition (iron abundance, transformed to
$Z$) when available, or suitable values from the literature or solar
metallicity in other cases. Initially we required the DLEB components
to be strictly coeval, i.e., the radii and effective temperatures
should be matched simultaneously for the same age, at the observed
masses. As the figure of merit for identifying the best fits we used a
simple $\chi^2$ statistic, and performed interpolations in age within
each track for higher accuracy.  We found that this brute-force grid
procedure often produced very poor fits or puzzling results such as
stars of similar masses being assigned very different \aov\ or
\amix\ parameters, or extreme values reaching the limits of our
grid. Similarly disappointing results were obtained when using the
radii along with the temperature ratios instead of the individual
temperatures, or when fitting only the radii, or when setting the
mixing length parameters of the hotter (radiative) components to
$\amix = 1.7$, near the solar-calibrated value, in combination with
the previous choices.  As we expect neither the observations nor the
models to be perfect, we later relaxed the condition of strict
coevality and allowed the component ages to differ by up to 5\%. This
improved the situation in some cases, but many still produced bad
matches to the measurements, suggesting perhaps a problem with the
abundances.

\begin{figure*}
\centering
\includegraphics[width=17.5cm]{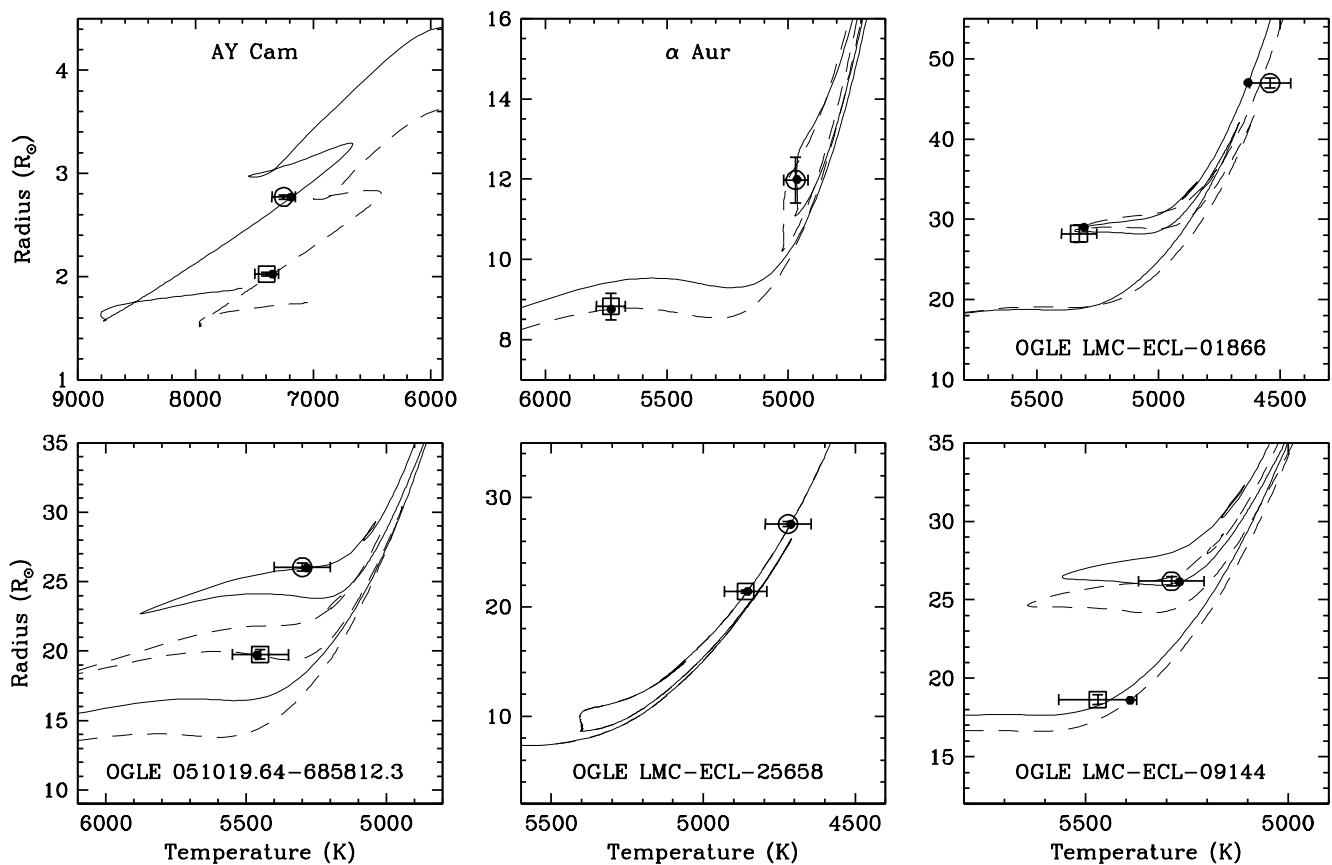}
\caption{Sample best fits to six of our binaries in the $R$
  vs.\ $T_{\rm eff}$ diagram. Evolutionary tracks and the observations
  for the primary in each system are represented with solid lines and
  open circles, while dashed lines and open squares are used for the
  secondary. Small dots mark the best-fit location on each track, and
  are always within the measurement uncertainties. \label{fig:rteff} }
\end{figure*}

Adding $Z$ as an extra dimension in our grids was considered too
computationally expensive given the number of binaries in our sample.
Therefore, using the above results as a guide, we performed the
adjustments manually system by system, varying $Z$ (assumed to be the
same for the two stars) along with \aov\ and \amix. As before, we
allowed age differences up to 5\%, though in most cases we found that
they came out much closer than this. Satisfactory fits were obtained
for the vast majority of our DLEBs, though some of them had a
preference for $Z$ values that were not insignificantly different from
those assumed initially, indicating that composition is indeed a
critical ingredient for the fits.  We discuss this issue in more
detail below.  Figure~\ref{fig:rteff} shows a few representative
examples of the fits.

\section{Results}
\label{sec:results}

As can be inferred from the properties in Table~\ref{tab:sample}, or
more directly from Figure~\ref{fig:rteff}, some of our systems are in
very rapid phases of evolution where ambiguities can sometimes occur
as to the location of the components in the H-R (or radius
vs.\ temperature) diagram. An advantage of our manual adjustments over
the blind grid fits is that they allow us to avoid inconsistencies
such as the primary appearing near the secondary in the $R$
vs.\ $T_{\rm eff}$ diagram but in a less evolved state, because of
observational errors.  Additionally, a few of our binaries have
measured primary and secondary masses that are close or
indistinguishable within the uncertainties, but they have very
different radii and temperatures. Manual inspection allowed us to
identify several cases where the published identities of the
components are reversed (see Table~\ref{tab:sample}).

Our results for the best-fit overshooting parameters, mixing length
parameters, bulk abundances, and mean ages are presented in
Table~\ref{tab:fitted}, where the systems are listed in the same order
as in Table~\ref{tab:sample}.  Approximate uncertainties for these
values were estimated through experiments in which we varied \aov,
\amix, or $Z$ and examined the goodness of fit, always requiring the
component ages to be within about 5\% of each other. The uncertainties
are strong functions of the evolutionary state, as this determines how
sensitive the predicted properties ($R$, $T_{\rm eff}$) are to the
fitted parameters. Typical theoretical errors for \aov\ were found to
be $\pm 0.03$ for evolved stars (giants), and $\pm 0.04$ for
main-sequence stars, in which overshooting has less of an
influence. Mixing length parameters have larger uncertainties of $\pm
0.20$ for each star. Abundance errors are typically 25--30\% in
$Z$. We note also that the $Z$ errors are very strongly correlated
with the uncertainties in \aov\ and \amix.

\begin{table*}
\caption{Fitted parameters for our sample of DLEBs.\label{tab:fitted}}
\centering
\begin{tabular}{l c c c c c c}
\hline\hline
     & \multicolumn{2}{c}{Primary} & \multicolumn{2}{c}{Secondary} &&  \\
Name & \aov\ & \amix\ & \aov\ & \amix\ & $Z$ & Mean age (Myr) \\
\noalign{\vskip 1pt}\hline\noalign{\vskip 2pt}
SMC-108.1-14904         &  0.250  &  2.10  &  0.240  &  1.95  &   0.0020  &   133    \\  
OGLE-LMC-ECL-CEP-0227   &  0.220  &  1.90  &  0.232  &  2.05  &   0.0018  &   142    \\  
OGLE-LMC-ECL-06575      &  0.230  &  1.90  &  0.230  &  1.92  &   0.0060  &   161    \\  
OGLE-LMC-ECL-CEP-2532   &  0.228  &  1.81  &  0.220  &  1.60  &   0.0017  &   164    \\  %
LMC-562.05-9009         &  0.170  &  1.95  &  0.165  &  1.95  &   0.0020  &   196    \\  
$\chi^2$ Hya            &  0.200  &  1.80  &  0.200  &  1.80  &   0.0110  &   214*   \\  %
OGLE-LMC-ECL-26122      &  0.200  &  1.90  &  0.200  &  1.75  &   0.0050  &   214    \\  
OGLE-LMC-ECL-01866      &  0.185  &  1.70  &  0.185  &  1.78  &   0.0045  &   224    \\  %
OGLE-SMC-113.3-4007     &  0.250  &  2.00  &  0.250  &  1.90  &   0.0030  &   230    \\  
OGLE-LMC-ECL-10567      &  0.200  &  1.70  &  0.200  &  1.75  &   0.0040  &   255    \\  
OGLE-LMC-ECL-09144      &  0.250  &  2.10  &  0.250  &  2.10  &   0.0030  &   256    \\  
OGLE-051019.64-685812.3 &  0.230  &  1.95  &  0.150  &  2.00  &   0.0040  &   278    \\  
OGLE-LMC-ECL-09660      &  0.200  &  1.75  &  0.250  &  1.75  &   0.0030  &   346    \\  
SMC-101.8-14077         &  0.200  &  1.75  &  0.200  &  1.90  &   0.0015  &   372    \\  
$\alpha$ Aur            &  0.230  &  1.54  &  0.230  &  1.60  &   0.0100  &   569    \\  
WX Cep                  &  0.170  &  1.80  &  0.170  &  1.80  &   0.0200  &   527    \\  
V1031 Ori               &  0.205  &  1.80  &  0.160  &  1.80  &   0.0200  &   616    \\  
V364 Lac                &  0.210  &  1.80  &  0.210  &  1.80  &   0.0200  &   622    \\  
SZ Cen                  &  0.210  &  1.80  &  0.210  &  1.80  &   0.0090  &   665    \\  
YZ Cas                  &  0.190  &  1.80  &  0.030  &  1.80  &   0.0100  &   645*   \\  
OGLE-LMC-ECL-25658      &  0.194  &  1.81  &  0.194  &  1.81  &   0.0043  &   819    \\  
V885 Cyg                &  0.190  &  1.80  &  0.190  &  1.80  &   0.0130  &   736*   \\  
AI Hya                  &  0.208  &  1.80  &  0.208  &  1.80  &   0.0300  &   955    \\  
VV Crv                  &  0.180  &  1.80  &  0.080  &  1.80  &   0.0350  &  1052*   \\  
AY Cam                  &  0.160  &  1.80  &  0.160  &  1.80  &   0.0200  &  1071    \\  
HY Vir                  &  0.075  &  1.80  &  0.020  &  1.80  &   0.0300  &  1363    \\  
SMC-130.5-04296         &  0.180  &  1.90  &  0.100  &  2.05  &   0.0020  &  1020    \\  
OGLE-LMC-ECL-03160      &  0.100  &  1.74  &  0.100  &  1.82  &   0.0025  &  1080    \\  
EI Cep                  &  0.140  &  1.80  &  0.140  &  1.80  &   0.0150  &  1405    \\  
SMC-126.1-00210         &  0.100  &  1.92  &  0.100  &  1.88  &   0.0025  &  1373    \\  
HD 187669               &  0.100  &  1.68  &  0.100  &  1.70  &   0.0100  &  2493    \\  
OGLE-LMC-ECL-15260      &  0.050  &  1.90  &  0.050  &  1.80  &   0.0030  &  2299    \\  
AI Phe                  &  0.040  &  1.70  &  0.000  &  1.85  &   0.0120  &  4995    \\  
\hline
\end{tabular}
\tablefoot{Systems marked with asterisks have components with age differences greater than 5\% (see text).}
\end{table*}

\begin{figure}
\centering
\includegraphics[width=8.5cm]{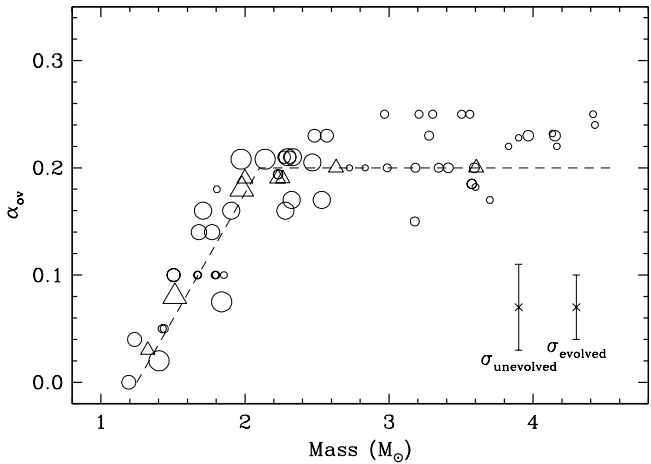}

\caption{Semi-empirical determination of the overshooting parameter
  \aov\ as a function stellar mass for the stars in the 33 systems of
  our sample. Primary and secondary components are plotted
  together. The size of the points is proportional to $Z$, the bulk
  composition (metal-poor stars have smaller symbols). Stars indicated
  with triangles are ones in which the inferred age difference within
  the binary exceeds our 5\% tolerance (see text), but which give
  otherwise acceptable fits to the observations. Typical error bars
  for dwarfs and giants are shown on the bottom right.
 \label{fig:overshooting}}

\end{figure}

A graphical representation of \aov\ as a function of stellar mass
appears in Figure~\ref{fig:overshooting}, which shows the primary and
secondary components together. Typical uncertainties described above
are indicated in the lower-right corner, for evolved and unevolved
stars.  A clear pattern is seen in \aov\ with a significant rise up to
about $2~M_{\sun}$, followed by little or no change beyond this
mass. The size of the symbols has been drawn proportional to $Z$.
While this conveys in a more visual way the fact that the higher mass
stars in our sample are all metal-poor (and belong to the LMC or SMC,
as seen in Table~\ref{tab:sample}) and that most low-mass stars have
higher abundances (they are typically solar neighborhood field stars),
we see no significant difference in \aov\ with $Z$ at a given mass, at
least in the present sample. The straight dashed lines in the figure
delineate the apparent trend with mass. Four systems (V885~Cyg,
$\chi^2$~Hya, VV~Crv, and YZ~Cas) have best fits yielding component
ages different by more than 5\%. The age discrepancies are,
respectively, 7\%, 15\%, 15\%, and 29\%. The most egregious case of
YZ~Cas has been notoriously difficult to fit in the past, and remains
so despite the recent observational efforts by \cite{Pavlovski:14},
who redetermined the masses, radii, temperatures, and chemical
composition.  These four peculiar systems are represented with
triangles in Figure~\ref{fig:overshooting}, but otherwise seem to
follow the same trend as the other binaries. Also, as noted earlier,
the primary components of OGLE-LMC-ECL-CEP-0227,
OGLE-LMC-ECL-CEP-2532, and LMC-562.05-9009 are classical Cepheids (the
first and last are fundamental-mode pulsators, and the other is a
first-overtone pulsator), although again their overshooting parameters
do not seem out of the ordinary for their mass.

\begin{figure}
\centering
\includegraphics[width=8cm]{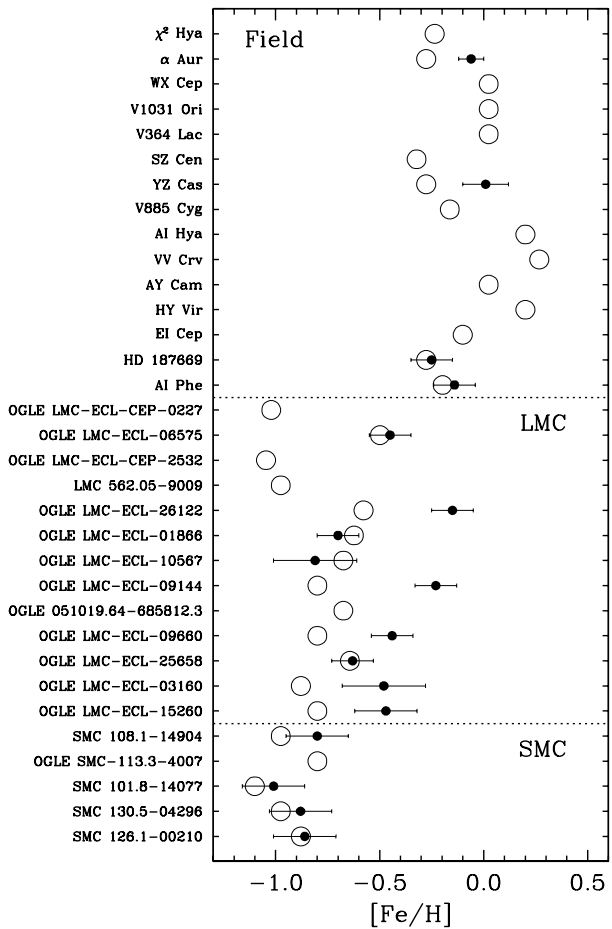}

\caption{Comparison between the available measured metallicities for
  the DLEBs in our sample (points with error bars) and the best-fit
  metallicities from our modeling (open circles), converted from $Z$
  to [Fe/H] as described in the text. Within each subgroup the binary
  systems are sorted as in Table~\ref{tab:sample}, by decreasing
  primary mass.
 \label{fig:z}}

\end{figure}

As mentioned before, our manual fits to the observations yield $Z$
values that are often rather different from the corresponding
spectroscopically measured composition, for systems in which this is
available.  We show this in Figure~\ref{fig:z}, where we have
transformed the fitted values of $Z$ to the corresponding [Fe/H]
measure assuming solar-scaled abundances and $Z_{\sun} = 0.0189$, to
enable a more direct comparison in the observational plane. We have
also segregated the systems from the LMC, the SMC, and the field, as
stars within these groups tend to have similar measured abundances.
Only about half of the systems in our sample have a measured
composition (see Table~\ref{tab:sample}).

Our fitted $Z$ values (open circles) are often systematically lower
than the corresponding spectroscopic [Fe/H] values (most noticeably in
the LMC systems), sometimes by as much as a factor of two or three.
In many cases this is well beyond the formal observational
uncertainties. The reasons for this are unclear.  Unrecognized biases
in the measured [Fe/H] values or in the temperatures (which would need
to be too hot by some 150 or 200~K, depending on the system) could
explain these differences, but they would have to be systematic in
nature as no instances of significantly overestimated best-fit $Z$
values were found. We also investigated the impact of the abundance
scale in our models. Tests showed that changing the reference solar
abundances from those of \cite{Grevesse:98} to those of
\cite{Asplund:09} does not make a significant difference in the
corresponding fitted $Z$ values. It is also possible that some of our
objects (particularly the more metal-poor ones) are enhanced in
$\alpha$ elements so that $[\alpha/{\rm Fe}] > 0$, which would change
the way we convert the model-fitted $Z$ values to the inferred
model-fitted [Fe/H] values shown in Figure~\ref{fig:z}
\citep[e.g.,][]{DAntona:13}. However, this effect goes in the wrong
direction to explain what we see.  Additional tests indicate that it
is in fact possible to match the measured [Fe/H] values but only at
the price of increasing \amix\ to values significantly larger than
2.0, and even approaching 3.0 in some cases. However, there is no
precedent for such large values of the mixing length parameter.
Finally, it may also be that the discrepancy illustrated in the figure
has to do with one or more of the physical ingredients in the
models. A comparison of our fitted $Z$ values with those fitted by
others for systems where this check is possible gave mixed results: in
some cases there is agreement, but in others our values are either
larger or smaller.

A closer look at the results in Table~\ref{tab:fitted} seems to
indicate a slight tendency for the fitted \amix\ values to be larger
in the LMC and SMC compared to the field binaries. The Magellanic
Cloud populations are more metal-poor than the field, on average,
suggesting there may be a dependence of the mixing length parameter on
metallicity. We note, however, that this apparent trend of larger
\amix\ values at lower abundances is opposite to that reported by
\cite{Bonaca:12}, who based their study on measured seismic
oscillation frequencies in a sample of single dwarf and subgiant stars
observed by the {\it Kepler} spacecraft. There are significant
differences between their study and ours that make it difficult to
pinpoint the reason(s) for the disagreement. For example, the stellar
masses and radii in our analysis are strictly empirical
(model-independent) as they were derived from DLEBs, whereas those of
\cite{Bonaca:12} were inferred from an initial fit to stellar
evolution models, and used again in a subsequent fit to infer
\amix. Additionally, the $T_{\rm eff}$, [Fe/H], and $\log g$ ranges in
our sample are all considerably larger than in the asteroseismic
sample. In particular, \cite{Bonaca:12} have only one star with $\log
g < 3.6$, while our sample contains many. In fact, it is quite
possible that the tentative differences we see in \amix\ between the
field binaries and the LMC/SMC binaries are related in part to
evolutionary effects, as all of our systems in the LMC/SMC are red
giants, in addition to being metal-poor.  Deficiencies in the stellar
evolution models that may influence the fitted \amix\ values cannot be
ruled out at the present time (particularly in light of the
fitted/measured abundance discrepancy discussed above), and the same
holds for potential systematic errors in the measurement of $T_{\rm
  eff}$ and/or [Fe/H], as noted earlier. For these reasons it may be
premature to claim a firm detection of a correlation between
\amix\ and [Fe/H] or $\log g$, though the hints we see certainly
warrant further investigation.

\section{Discussion}
\label{sec:discussion}

The dependence of the overshooting parameter \aov\ on stellar mass
presented in Figure~\ref{fig:overshooting} is much more clearly
established than in previous work, including that of \cite{Ribas:00}
and \cite{Claret:07}, which suffered from a lack of suitable binary
systems. Not only is the overall size of our binary list significantly
larger, but the critical 1.2--2.0~$M_{\sun}$ range in which
\aov\ ramps up from zero to an apparent maximum is much better sampled
as well. The negative results of some of the more recent DLEB studies
that did not see any mass dependence of \aov\ can generally be
understood in terms of the binary systems they used. For instance, of
the four objects examined by \cite{Meng:14}, some are similar
regarding their component masses and this makes the sample effectively
smaller. Furthermore, overshooting is assumed to be the same for the
two components in each system, which would make it more difficult to
detect any real change as a function of mass, especially since those
binaries all have mass ratios significantly different from unity. For
reasons that are not understood, unequal binaries such as these tend
to be problematic to model, as we ourselves have found for
$\chi^2$~Hya, VV~Crv, and YZ~Cas, three of the four objects in the
\cite{Meng:14} sample. A similar modeling challenge was mentioned
earlier for the unequal-mass binary V380~Cyg, which weighed heavily in
the results of \cite{Ribas:00}.  The stars analyzed by
\cite{Stancliffe:15}, on the other hand, cover a fairly wide range of
masses, although most are clustered around 2~$M_{\sun}$ and several
are relatively unevolved and are therefore less sensitive to the
effects of overshooting.

An interesting feature of our sample is that it spans a wide range of
measured metallicities equivalent to a full factor of ten, from [Fe/H]
of $-1.01$ to $+0.01$. Despite this, we are unable to discern any
dependence of \aov\ on $Z$ at a given mass, suggesting perhaps that
the precision of our \aov\ determinations would need to be
considerably better to detect such an effect, if it exists, or that
the sample needs to be even larger.

\begin{figure}
\centering
\includegraphics[width=8.5cm]{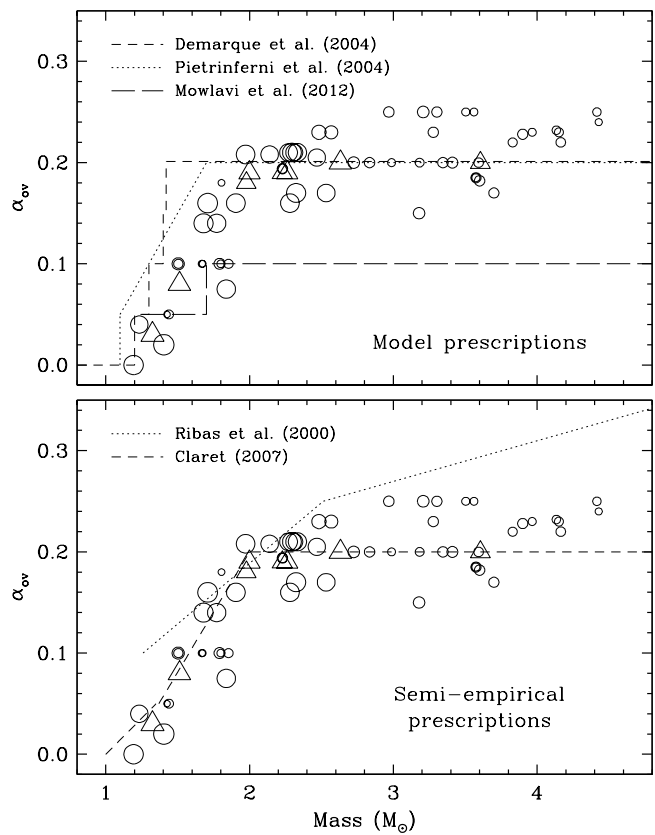}

\caption{Our semi-empirical determinations of \aov\ as a function of
  stellar mass, compared with published prescriptions. Symbols are as
  in Figure~\ref{fig:overshooting}, with sizes drawn here proportional
  to the surface gravity $\log g$ of each star. \emph{Top:} Mass
  dependence of \aov\ adopted in some recent grids of stellar
  evolution models.  \emph{Bottom:} Comparison with previous
  semi-empirical relationships by \cite{Ribas:00} and
  \cite{Claret:07}. \label{fig:prescriptions} }

\end{figure}

Many current stellar evolution models implement overshooting with a
built-in dependence on stellar mass (and sometimes metallicity), even
though the exact shape of that dependence has so far been poorly
constrained by observations, except in the general sense that
overshooting is irrelevant below 1.1--1.3~$M_{\sun}$, then grows, and
seems to level off at some larger mass. The top panel of
Figure~\ref{fig:prescriptions} shows how some of those model
prescriptions fare against our new results. In the \cite{Demarque:04}
overshooting implementation the rise in \aov\ from zero is steeper
than indicated by our measurements, but eventually reaches a similar
maximum around 0.2. The \cite{Pietrinferni:04} formula appears to have
the right slope, but the rise starts and ends at lower masses.  The
models of \cite{Mowlavi:12} use a recipe for \aov\ that grows to only
half the peak value indicated by our analysis.  It is worth mentioning
here that our measurements of \aov\ are semi-empirical in nature
because they are based on the observed properties of binary systems
but they also depend on models, specifically the Granada series of
\cite{Claret:04, Claret:12}. Although this may suggest our results
have limited applicability, the key physical ingredients in most
modern stellar evolution codes (standard mixing-length approximation,
radiative opacities, etc.) are rather similar, so we expect our
conclusions to be valid for other models as well.

Previous semi-empirical results are compared with ours in the lower
panel of Figure~\ref{fig:prescriptions}. As noted earlier, the
\cite{Ribas:00} mass dependence of \aov\ is steeper than indicated by
our measurements, while that of \cite{Claret:07}, which was based on a
smaller sample than ours and had much larger uncertainties in \aov, is
fairly close to the trend indicated by our results, except possibly at
the very lowest masses, where that study had only one star (the
secondary of YZ~Cas). This study also suggested a slight increase in
\aov\ for stars more massive than 10~$M_{\sun}$, a regime our sample
does not address.

An assumption that is implicit in the present work, and indeed in all
stellar evolution codes we are aware of, is that \aov\ does not depend
on the evolutionary state of the star at a given mass, i.e., it does
not vary with time. With the usual expression $d_{\rm ov} = \aov\ H_p$
for the distance traveled by convective cells above the boundary of
the core, it is clear that $d_{\rm ov}$ changes as a star evolves
because $H_p$ does. But \aov\ could well vary independently in some
fashion \citep[as speculated, e.g., by][]{Torres:14}. Our measurements
allow a first look into this issue.  In Figure~\ref{fig:prescriptions}
we have represented the \aov\ measurements with symbols whose size is
proportional to the surface gravity of the star ($\log g$), in order
to more easily distinguish dwarfs from giants. While most stars on the
rising part of the \aov\ vs.\ mass trend ($M < 2~M_{\sun}$) are
main-sequence stars, some are low mass giants (smaller symbols), and
there seems to be no significant difference between the overshooting
parameters of dwarfs and giants of similar mass. The behavior at
higher masses cannot be addressed with the present sample for lack of
sufficiently evolved main-sequence binaries with $M > 2~M_{\sun}$.

\begin{figure}
\centering
\includegraphics[width=8.5cm]{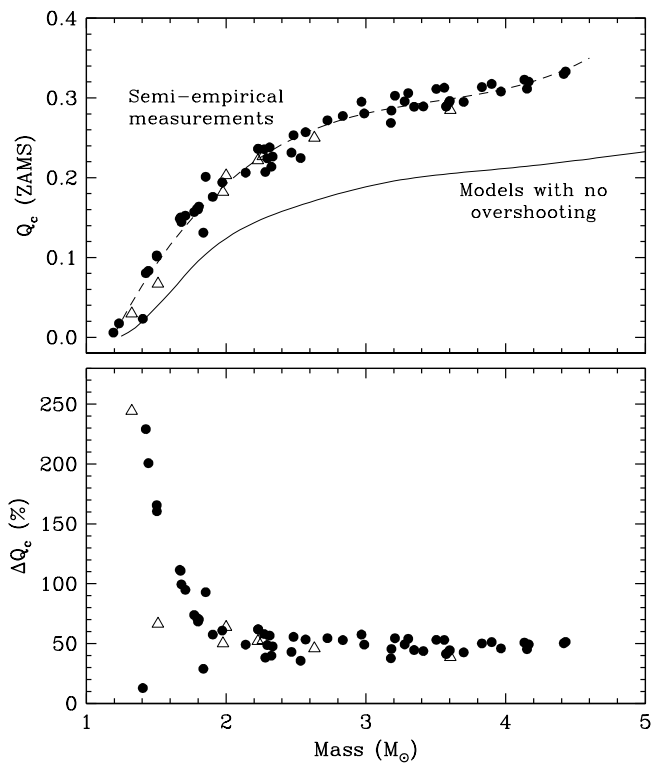}
\caption{\emph{Top:} Semi-empirical values $Q_c$ of the convective
  core mass at the ZAMS (normalized to the total mass) for each of the
  stars in our sample. Triangles represent components of systems with
  age differences exceeding 5\%. The dashed line is a third-order
  polynomial fit drawn to guide the eye. The solid line shows the
  trend of $Q_c$ with mass for solar-metallicity ZAMS models having
  $\aov = 0$, for reference. \emph{Bottom:} Fractional increase in
  $Q_c$ over that indicated by the solid line in the top panel,
  expressed as a percentage.
 \label{fig:qc} }
\end{figure}

The clear evidence from our semi-empirical measurements that the
influence of overshooting initially rises as the mass increases from
about 1.2~$M_{\sun}$ carries some interest in itself from the
theoretical point of view, as it must contain quantitative information
about the implied growth of the convective core. We investigated this
using the same best-fit models for the stars in our sample from which
we obtained \aov. As the size (mass) of the core also changes with
time as stars evolve, we have chosen to eliminate the time dependence
by extracting from each model used to generate
Figure~\ref{fig:overshooting} the mass of the convective core at the
ZAMS, and we then normalized it to the total mass of the star.  The
results for the fractional core mass $Q_c$ obtained in this way are
shown in the top panel of Figure~\ref{fig:qc}, displayed as a function
of stellar mass. For reference we have added a solid curve
representing the predicted change in $Q_c$ in the absence of
overshooting, also at the ZAMS. While this last curve clearly
indicates, as expected, that the core mass grows with stellar mass
even without overshooting, the increase is considerably more
pronounced with overshooting, and our semi-empirical measurements of
$Q_c$ allow us to quantify the degree to which this is the case. The
lower panel displays the fractional increase in $Q_c$ as a function of
stellar mass, and indicates that for stars beyond about 2~$M_{\sun}$
it converges toward an enlargement of about 50\% above the core mass
that would result in the absence of overshooting.  This differential
increase raises additional questions of interest. What is the maximum
possible size of the convective core for a given stellar mass? How
does \aov\ influence the size of the core at different stellar masses?
With relatively simple arguments and the use of the virial theorem it
can be shown that there is in fact an upper limit to the size of the
mixed core, implying a limit to \aov, as we see from our
measurements. Thus, it is possible to understand the general features
of Figure~\ref{fig:overshooting}, at least over the mass range
explored here.  The details of these calculations are gathered in the
Appendix for the interested reader.

\section{Conclusions}
\label{sec:conclusions}

We have used the measured masses, radii, and effective temperatures of
more than 30 carefully selected double-lined eclipsing binary systems
to infer semi-empirical values of \aov\ for each of the components by
comparison with stellar evolution models. Importantly, the sample
includes a substantial number of highly evolved systems (red giants,
mostly in the LMC and SMC) that are more sensitive to the effects of
overshooting. This significantly larger and more suitable sample
compared to previous studies has allowed us to calibrate the
dependence of overshooting on stellar mass. This dependence has
traditionally been assumed to be present, and has been implemented in
various ways in current stellar evolution models, but until now it has
not been well constrained by observations. We find a clear and fairly
linear increase in \aov\ beginning at about 1.2~$M_{\sun}$ and
reaching $\aov \sim 0.2$ around 2.0~$M_{\sun}$, with little change
beyond this mass up to the limit of our sample (4.4~$M_{\sun}$). This
trend is similar to, but much better defined than that proposed by
\cite{Claret:07}, and differs in various ways from prescriptions
currently used in other model sets that adopt the same formulation for
the extension of the convective core as $d_{\rm ov} = \aov\ H_p$. Our
results may serve as a guide for future implementations of
overshooting in model grids. We also find no significant variance in
\aov\ for giants and dwarfs at a given mass, suggesting \aov\ does not
depend very strongly on evolutionary state. Three of our LMC systems
are Cepheids, though no significant difference is found in their
\aov\ values either compared to other stars of similar mass.  The main
features of the \aov\ vs.\ mass trend, which is the main result of
this work, can be understood by simple physical arguments as laid out
in the Appendix.

We also made use of the same best-fit models for our 33 binary systems
to calibrate the extent of the convective core as a function of
stellar mass. We find that for stars more massive than about
2~$M_{\sun}$ the growth of the fractional core mass $Q_c$ converges to
a level of about 50\% above the values predicted by models without
overshooting.

All of our binary systems yield satisfactory fits when compared with
the models, with the exception of four (V885~Cyg, $\chi^2$~Hya,
VV~Crv, and YZ~Cas) in which the component ages differ by more than
5\% (particularly YZ~Cas), although their other properties are well
matched. The last three of these have mass ratios appreciably
different from unity, and the fact that another similarly unequal
system (V380~Cyg) has also been difficult to model in past studies by
others suggests either unrecognized measurement errors in such
systems, or some other problem that has yet to be identified, perhaps
related to their origin.

One half of the binaries in our sample have a spectroscopically
measured [Fe/H] abundance in the literature. In about half of those
cases we find curious systematic differences between the measured
composition and the $Z$ values from our best fits, after conversion to
the [Fe/H] scale: the fitted values tend to be smaller, often
significantly so. Most of these systems belong to the LMC. There are
no examples with opposite discrepancies of much significance, and the
reasons for this are not understood.  In some cases other authors have
found similar deviations, though this has not been emphasized.  There
are also hints in our sample that the fitted \amix\ values may be
higher for stars that are more metal-poor and/or more evolved, though
this may be related to the $Z$ differences just noted, and needs to be
investigated further.

Although much larger than previous lists of binaries used to calibrate
overshooting, our sample is still limited in that it contains no
systems with component masses beyond 4.4~$M_{\sun}$.  Therefore, we
are unable to verify claims by others about possible changes in
\aov\ for more massive stars.  While our objects do cover a range of
about a factor of ten in metal abundance, and we find no significant
dependence of \aov\ on $Z$ within our measurement uncertainties, a
larger metallicity range is desirable to confirm this conclusion, and
perhaps track down the deviations mentioned above.

\begin{acknowledgements}
We thank the anonymous referee for helpful comments on the original
manuscript.  The Spanish MEC (AYA2015-71718-R) is gratefully
acknowledged for its support during the development of this work. GT
acknowledges partial support from the NSF through grant
AST-1509375. This research has made use of the SIMBAD database,
operated at the CDS, Strasbourg, France, and of NASA's Astrophysics
Data System Abstract Service.
\end{acknowledgements}

\begin{appendix}
\label{sec:appendix}

\section{A study of extra mixing using the virial theorem}

\begin{figure}[b]
\centering
\includegraphics[width=8.5cm]{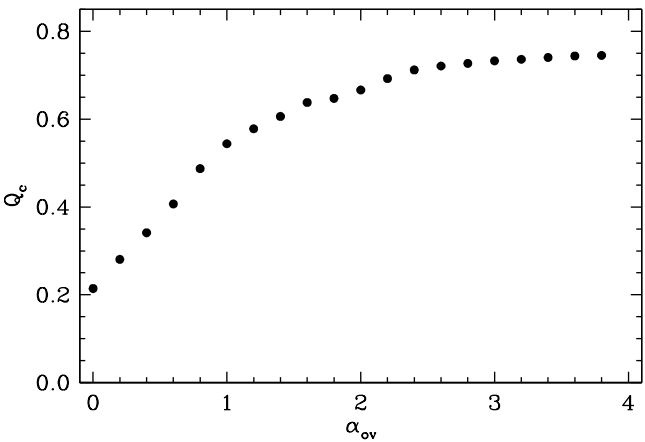}
\caption{Dependence of the fractional mass of the mixed core as a
  function of the overshooting parameter \aov, for models with
  $3~M_{\sun}$ and composition $X = 0.751$ and $Z = 0.003$. Note the
  linear behavior of $Q_c$ for $\aov \le 1.0$. \label{fig:A1} }
\end{figure}

In order to investigate the consequences of extra mixing on the
internal physical conditions of a star, we have selected a model of a
$3~M_{\sun}$ star with composition $X = 0.751$ and $Z = 0.003$ that is
representative of our observational sample. A first point of interest
is the issue of how the extra mixing depends on the \aov\ parameter,
defined as in the main text by $d_{\rm ov} = \aov H_p$, where $H_p$ is
the pressure scale height. Figure~\ref{fig:A1} shows the behavior of
the fractional mixed core mass $Q_c = M_c/M$ as a function of \aov\ at
the ZAMS, in which $M_c$ is the mass of the convective core and $M$
the total mass of the star. It can be seen that up to $\aov \approx
1.0$ the increase in $Q_c$ is essentially linear, and tests reveal
that the slope $d Q_c/d\aov \approx 0.3$ is almost independent of
stellar mass. For larger \aov\ values the figure suggests that a limit
to the size of the mixed core is eventually reached, such that for
\aov\ larger than about 2.5 the fractional core mass $Q_c$ is
practically independent of the overshooting parameter. For a closer
look into this limit we make use here of the virial theorem, a very
useful but frequently overlooked analytical tool for stellar physics.

The virial theorem for a star may be written as
\begin{eqnarray}
\label{eq:A1}
\zeta E_i + \Omega = 0,
\end{eqnarray}
in which $E_i =\int_{0}^{M} u~dm$ is the total internal energy,
$\Omega = -G\int_{0}^{M} m/r~dm \equiv -\alpha {GM^2/R}$ is the
gravitational potential energy, and $R$ is the stellar radius. In
addition, for an ideal gas we have $u = c_v T$ and $\zeta u = 3 P/\rho
= 3 (\gamma-1)$, and $\gamma = c_P/c_v$, where $u$ is the specific
internal energy, $P$ the pressure, $T$ the temperature, and $c_v$ and
$c_P$ are the specific heats at constant volume and constant pressure.

However, equation~\ref{eq:A1} derives from the hydrostatic
differential equation assuming that the pressure at the boundary
vanishes. In the more general case we may write the virial for the
mixed core as
\begin{eqnarray}
G\int_{0}^{M_c} \frac{m}{r}~dm = 3 \int_{0}^{M_c} \frac{P}{\rho} dm - 4 \pi r^3 P_c, 
\end{eqnarray}
where $P_c$ denotes the pressure at the boundary of the
core. Equation~\ref{eq:A1} then becomes
\begin{eqnarray}
\label{eq:A3}
\zeta E_i + \Omega = 4 \pi r^3 P_c.
\end{eqnarray}
 
Equation~\ref{eq:A3} may be used to study the consequences of the
extension of the core due to extra mixing under the condition of
hydrostatic equilibrium. To proceed we will assume for simplicity a
non-degenerate ideal gas. As the extra mixing increases (corresponding
to larger adopted values of \aov\ in our framework), the pressure at
the border of the core will decrease. We may estimate the maximum
extent of the extra mixing (in terms of the radial coordinate and
pressure) by solving $dP/dr$ = 0, that is,
\begin{eqnarray}
\frac{dP}{dr} = \frac{\alpha G M_c^2}{4\pi r^5}  - 
\frac{{3\left(\frac{-\alpha G M_c^2}{r} - 
\frac{3 k M_c\overline{T_c}}{(\gamma - 1)\mu}   + 
\frac{3 \gamma k M_c \overline{T_c}}{(\gamma - 1)\mu}\right)}}{4 \pi r^4} = 0.
\end{eqnarray}
From this, the critical radius $r_{\rm crit}$ is given by 
\begin{eqnarray}
{r_{\rm crit}} = \frac{4 \alpha G M_c\mu}{9 k \overline{T_c}},
\end{eqnarray}
where $\mu$ is the mean molecular weight, $k$ is the Boltzmann
constant, and $\overline{T_c} = 1/M_c \int_{0}^{M_c}
T~dm$. Correspondingly, there is a critical value for the pressure at
the border of the mixed core, $P_{\rm crit}$, that can hydrostatically
support the weight of the envelope:
\begin{eqnarray}
\label{eq:A6}
{P_{\rm crit}}= \frac{2187 k^4\overline{T_c}^4}{1024\pi G^3\alpha^3 M_c^2 \mu^4}.
\end{eqnarray}

The expression above shows that $P_{\rm crit}$ decreases as the core
mass increases, as expected. On the other hand, the pressure at the
bottom of the envelope is $P_e \propto \overline{T_c}^4/M^2$.  The
condition of hydrostatic equilibrium at the interface implies that
\begin{eqnarray}
\label{eq:A7}
P_{\rm crit} \ge P_e.
\end{eqnarray}

\begin{figure}[ht!]
\centering
\includegraphics[width=8.5cm]{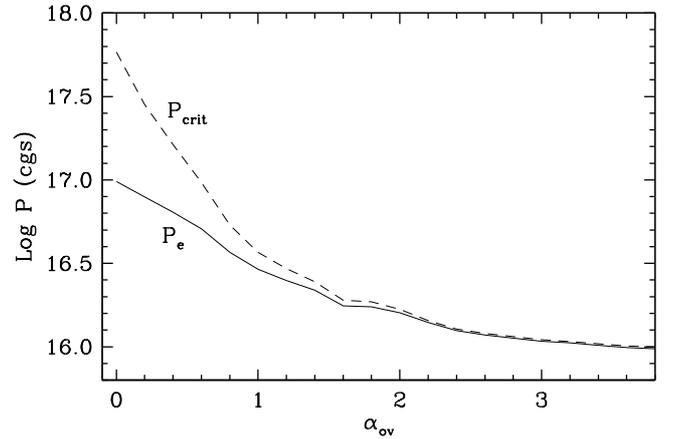}
\caption{The pressure on a logarithmic scale at the bottom of the
  envelope (solid line), compared with the critical pressure
  $P_{\rm crit}$ given by equation~\ref{eq:A6} (dashed line), as a
  function of the overshooting parameter \aov.  The models are the
  same as those in Figure~\ref{fig:A1}.\label{fig:A2} }
\end{figure}

The pressure at the bottom of the envelope can be extracted from the
models shown in Figure~\ref{fig:A1}, and compared with $P_{\rm crit}$.
This comparison appears in Figure~\ref{fig:A2}. Despite the
simplicity of the assumptions adopted for our use of virial theorem
(ideal non-degenerate gas, no radiation pressure, non-rotating models,
etc.), the results in Figure~\ref{fig:A2} are in good agreement with
those shown in Figure~\ref{fig:A1}: the curves in Figure~\ref{fig:A2}
meet at approximately the same value of \aov\ at which the trend in
Figure~\ref{fig:A1} levels off.  Applying equation~\ref{eq:A7} to
Figure~\ref{fig:A2} we may infer a limit to the size of the
mixed core (for the present case, $Q_{c,\rm max} \approx 0.75$) and
consequently a critical value for \aov\ around 2.0--2.5, although
given our assumptions and simplifications, this limit value could be
as small as $\aov \approx 1.5$. Mutatis mutandis, we expect the same
should occur for other masses typical of our observational sample. The
critical values of the radius and pressure are also influenced by the
initial chemical composition, not only through changes in the internal
structure (for example, the convective core for a $Z = 0.02$ model is
slightly smaller than for one with $Z = 0.003$), but also directly in
the calculation of $P_{\rm crit}$ (see above).

\end{appendix}
\newpage

\longtab[1]{
\begin{longtable}{llllcc}
\caption{Binaries systems in our sample.\label{tab:sample}} \\
\hline\hline\noalign{\vskip 2pt}
Name & Mass ($M_{\sun}$) & Radius ($R_{\sun}$) & $T_{\rm eff}$ (K) & [Fe/H] & Source \\
\noalign{\vskip 1pt}\hline\noalign{\vskip 3pt}
\endfirsthead
\caption{Continued.} \\
\hline\hline\noalign{\vskip 2pt}
Name & Mass ($M_{\sun}$) & Radius ($R_{\sun}$) & $T_{\rm eff}$ (K) & [Fe/H] & Source \\
\noalign{\vskip 1pt}\hline\noalign{\vskip 3pt}
\endhead
\hline
\endfoot
SMC-108.1-14904         &   4.416~$\pm$~0.041   &  46.95~$\pm$~0.53   &  5675~$\pm$~105  & $-$0.80~$\pm$~0.15 &  1  \\
                        &   4.429~$\pm$~0.037   &  64.05~$\pm$~0.50   &  4955~$\pm$~90   &                    &     \\ [1ex]
OGLE-LMC-ECL-CEP-0227   &   4.165~$\pm$~0.032   &  34.92~$\pm$~0.34   &  6050~$\pm$~160  &                    &  2  \\
                        &  4.134~$\pm$~0.037    &  44.85~$\pm$~0.29   &  5120~$\pm$~130  &                    &     \\ [1ex]
OGLE-LMC-ECL-06575      &   4.152~$\pm$~0.030   &  39.79~$\pm$~1.35   &  4903~$\pm$~72   & $-$0.45~$\pm$~0.10 &  3  \\
                        &  3.966~$\pm$~0.032    &  49.35~$\pm$~1.45   &  4681~$\pm$~77   &                    &     \\ [1ex]
OGLE-LMC-ECL-CEP-2532   &    3.90~$\pm$~0.10    &  28.95~$\pm$~1.4    &  6345~$\pm$~150  &                    &  4  \\
                        &   3.83~$\pm$~0.10     &   37.7~$\pm$~1.7    &  4800~$\pm$~220  &                    &     \\ [1ex]
LMC-562.05-9009         &    3.70~$\pm$~0.03    &   28.6~$\pm$~0.2    &  6030~$\pm$~150: &                    &  5  \\
                        &   3.60~$\pm$~0.03     &   26.6~$\pm$~0.2    &  6030~$\pm$~150: &                    &     \\ [1ex]
$\chi^2$ Hya            &   3.605~$\pm$~0.078   &  4.390~$\pm$~0.039  & 11750~$\pm$~190  &                    &  6  \\
                        &   2.632~$\pm$~0.049   &  2.159~$\pm$~0.030  & 11100~$\pm$~230  &                    &     \\ [1ex]
OGLE-LMC-ECL-26122      &   3.593~$\pm$~0.055   &  32.71~$\pm$~0.51   &  4989~$\pm$~80   & $-$0.15~$\pm$~0.10 &  3  \\
                        &  3.411~$\pm$~0.047    &  22.99~$\pm$~0.48   &  4995~$\pm$~81   &                    &     \\ [1ex]
OGLE-LMC-ECL-01866      &  3.574~$\pm$~0.038    &  46.96~$\pm$~0.61   &  4541~$\pm$~85   & $-$0.70~$\pm$~0.10 &  3* \\
                        &   3.575~$\pm$~0.028   &  28.20~$\pm$~1.06   &  5327~$\pm$~72   &                    &     \\ [1ex]
OGLE-SMC-113.3-4007     &   3.561~$\pm$~0.025   &   48.4~$\pm$~0.7    &  4813~$\pm$~100  &                    &  7*  \\
                        &  3.504~$\pm$~0.028    &   45.8~$\pm$~0.7    &  4800~$\pm$~100  &                    &     \\ [1ex]
OGLE-LMC-ECL-10567      &   3.345~$\pm$~0.040   &   25.6~$\pm$~1.6    &  5067~$\pm$~73   & $-$0.81~$\pm$~0.20 &  3  \\
                        &  3.183~$\pm$~0.038    &   36.0~$\pm$~2.0    &  4704~$\pm$~80   &                    &     \\ [1ex]
OGLE-LMC-ECL-09144      &   3.303~$\pm$~0.028   &  26.18~$\pm$~0.31   &  5288~$\pm$~81   & $-$0.23~$\pm$~0.10 &  3  \\
                        &  3.208~$\pm$~0.026    &  18.64~$\pm$~0.30   &  5470~$\pm$~96   &                    &     \\ [1ex]
OGLE-051019.64-685812.3 &   3.278~$\pm$~0.032   &  26.05~$\pm$~0.29   &  5300~$\pm$~100  &                    &  6  \\
                        &  3.179~$\pm$~0.029    &  19.76~$\pm$~0.34   &  5450~$\pm$~100  &                    &     \\ [1ex]
OGLE-LMC-ECL-09660      &   2.988~$\pm$~0.018   &  43.87~$\pm$~1.14   &  4677~$\pm$~75   & $-$0.44~$\pm$~0.10 &  3*  \\
                        &  2.969~$\pm$~0.020    &  23.75~$\pm$~0.66   &  5352~$\pm$~70   &                    &     \\ [1ex]
SMC-101.8-14077         &   2.835~$\pm$~0.055   &  23.86~$\pm$~0.31   &  5170~$\pm$~90   & $-$1.01~$\pm$~0.15 &  1*  \\
                        &  2.725~$\pm$~0.034    &  17.90~$\pm$~0.50   &  5580~$\pm$~95   &                    &     \\ [1ex]
$\alpha$ Aur            &  2.5687~$\pm$~0.0074  &  11.98~$\pm$~0.57   &  4970~$\pm$~50   & $-$0.04~$\pm$~0.06 &  8  \\
                        & 2.4828~$\pm$~0.0067   &   8.83~$\pm$~0.33   &  5730~$\pm$~60   &                    &     \\ [1ex]
WX Cep                  &   2.533~$\pm$~0.050   &  3.996~$\pm$~0.030  &  8150~$\pm$~250  &                    &  6  \\
                        &  2.324~$\pm$~0.045    &  2.712~$\pm$~0.023  &  8900~$\pm$~250  &                    &     \\ [1ex]
V1031 Ori               &   2.468~$\pm$~0.018   &  4.323~$\pm$~0.034  &  7850~$\pm$~500  &                    &  6  \\
                        &  2.281~$\pm$~0.016    &  2.978~$\pm$~0.064  &  8400~$\pm$~500  &                    &     \\ [1ex]
V364 Lac                &   2.333~$\pm$~0.014   &  3.309~$\pm$~0.021  &  8250~$\pm$~150  &                    &  6  \\
                        &  2.295~$\pm$~0.024    &  2.986~$\pm$~0.020  &  8500~$\pm$~150  &                    &     \\ [1ex]
SZ Cen                  &   2.311~$\pm$~0.026   &  4.556~$\pm$~0.032  &  8100~$\pm$~300  &                    &  6  \\
                        &  2.272~$\pm$~0.021    &  3.626~$\pm$~0.026  &  8380~$\pm$~300  &                    &     \\ [1ex]
YZ Cas                  &   2.263~$\pm$~0.012   &  2.525~$\pm$~0.011  &  9520~$\pm$~120  & $+$0.01~$\pm$~0.11 &  9  \\
                        &  1.325~$\pm$~0.007    &  1.331~$\pm$~0.006  &  6880~$\pm$~240  &                    &     \\ [1ex]
OGLE-LMC-ECL-25658      &  2.230~$\pm$~0.019    &  27.57~$\pm$~0.24   &  4721~$\pm$~75   & $-$0.63~$\pm$~0.10 & 10* \\
                        &  2.229~$\pm$~0.019    &  21.41~$\pm$~0.15   &  4860~$\pm$~70   &                    &     \\ [1ex]
V885 Cyg                &   2.228~$\pm$~0.026   &  3.387~$\pm$~0.026  &  8150~$\pm$~150  &                    &  6  \\
                        &  2.000~$\pm$~0.029    &  2.346~$\pm$~0.017  &  8375~$\pm$~150  &                    &     \\ [1ex]
AI Hya                  &   2.140~$\pm$~0.038   &  3.916~$\pm$~0.031  &  6700~$\pm$~60   &                    &  6  \\
                        &  1.973~$\pm$~0.036    &  2.767~$\pm$~0.019  &  7100~$\pm$~65   &                    &     \\ [1ex]
VV Crv                  &   1.978~$\pm$~0.010   &  3.375~$\pm$~0.010  &  6500~$\pm$~200  &                    & 11 \\
                        &  1.513~$\pm$~0.008    &  1.650~$\pm$~0.008  &  6638~$\pm$~200  &                    &     \\ [1ex]
AY Cam                  &   1.905~$\pm$~0.040   &  2.772~$\pm$~0.020  &  7250~$\pm$~100  &                    &  6  \\
                        &  1.709~$\pm$~0.036    &  2.026~$\pm$~0.017  &  7395~$\pm$~100  &                    &     \\ [1ex]
HY Vir                  &   1.838~$\pm$~0.009   &  2.806~$\pm$~0.008  &  6850~$\pm$~130  &                    & 12  \\
                        &  1.404~$\pm$~0.006    &  1.519~$\pm$~0.008  &  6550~$\pm$~120  &                    &     \\ [1ex]
SMC-130.5-04296         &   1.805~$\pm$~0.027   &  46.00~$\pm$~0.35   &  4515~$\pm$~75   & $-$0.88~$\pm$~0.15 &  1*  \\
                        &  1.854~$\pm$~0.025    &  25.44~$\pm$~0.25   &  4912~$\pm$~80   &                    &     \\ [1ex]
OGLE-LMC-ECL-03160      &   1.792~$\pm$~0.027   &  16.36~$\pm$~1.06   &  4954~$\pm$~83   & $-$0.48~$\pm$~0.20 &  3  \\
                        &  1.799~$\pm$~0.028    &  37.42~$\pm$~0.52   &  4490~$\pm$~82   &                    &     \\ [1ex]
EI Cep                  &  1.7716~$\pm$~0.0066  &  2.897~$\pm$~0.048  &  6750~$\pm$~100  &                    &  6  \\
                        & 1.6801~$\pm$~0.0062   &  2.330~$\pm$~0.044  &  6950~$\pm$~100  &                    &     \\ [1ex]
SMC-126.1-00210         &   1.674~$\pm$~0.037   &  43.52~$\pm$~1.02   &  4480~$\pm$~70   & $-$0.86~$\pm$~0.15 &  1  \\
                        &  1.669~$\pm$~0.039    &  39.00~$\pm$~0.98   &  4510~$\pm$~70   &                    &     \\ [1ex]
HD 187669               &   1.505~$\pm$~0.004   &  22.62~$\pm$~0.50   &  4330~$\pm$~70   & $-$0.25~$\pm$~0.10 & 13*  \\
                        &  1.504~$\pm$~0.004    &  11.33~$\pm$~0.28   &  4650~$\pm$~80   &                    &     \\ [1ex]
OGLE-LMC-ECL-15260      &   1.440~$\pm$~0.024   &  23.51~$\pm$~0.69   &  4706~$\pm$~87   & $-$0.47~$\pm$~0.15 &  3  \\
                        &  1.426~$\pm$~0.022    &  42.17~$\pm$~0.33   &  4320~$\pm$~81   &                    &     \\ [1ex]
AI Phe                  &  1.2336~$\pm$~0.0045  &  2.932~$\pm$~0.048  &  5010~$\pm$~120  & $-$0.14~$\pm$~0.10 &  6  \\
                        & 1.1934~$\pm$~0.0041   &  1.818~$\pm$~0.024  &  6310~$\pm$~150  &                    &     \\
\end{longtable}
\tablebib{
1 - \cite{Graczyk:14}; 
2 - \cite{Pilecki:13}; 
3 - \cite{Pietrzynski:13}; 
4 - \cite{Pilecki:15}; 
5 - \cite{Gieren:15}; 
6 - \cite{Torres:10}; 
7 - \cite{Graczyk:12}; 
8 - \cite{Torres:15}; 
9 - \cite{Pavlovski:14}; 
10 - \cite{Elgueta:16};
11 - \cite{Fekel:13}; 
12 - \cite{Lacy:11}; 
13 - \cite{Helminiak:15}.
}

\tablefoot{
The first line for each system corresponds to the primary, and the
next to the secondary. Sources flagged with an asterisk indicate cases
where we have swapped the primary/secondary identification relative to
the original publication (see text).
Temperatures for LMC-562.05-9009 are listed as uncertain in the
original source. The [Fe/H] value adopted here for OGLE-LMC-ECL-25658
is the average of the individual estimates reported.
}
}


\begin{thebibliography}{}

\bibitem[Aerts (2013)]{Aerts:13} Aerts, C. 2013, in Setting a New
  Standard in the Analysis of Binary Stars, eds.\ K.\ Pavlovski,
  A.\ Tkachenko \& G.\ Torres. EAS Publications Series, Vol.\ 64,
  2013, pp.\ 323-330

\bibitem[Aerts (2015)]{Aerts:15} Aerts, C. 2015, in New Windows on
  Massive Stars: Asteroseismology, Interferometry, and
  Spectropolarimetry, Proceedings of the International Astronomical
  Union, IAU Symposium, Volume 307, pp.\ 154-164

\bibitem[Andersen(1991)]{Andersen:91} Andersen, J. 1991, \aapr, 3, 91

\bibitem[Andersen et al.(1990)]{Andersen:90} Andersen, J., Clausen,
  J.\ V., Nordstr\"om, B. 1990, \aap, 363, 33

\bibitem[Asplund et al.(2009)]{Asplund:09} Asplund, M., Grevesse, N.,
  Sauval, A.\ J., \& Scott, P. 2009, \araa, 47, 481

\bibitem[B\"ohm-Vitense(1958)]{Vitense:58} B\"ohm-Vitense, E. 1958,
  \zap, 46, 108

\bibitem[Bonaca et al.(2012)]{Bonaca:12} Bonaca, A., Tanner, J.\ D.,
  Basu, S.\ et al. 2012, \apjl, 755, L12

\bibitem[Bressan(1992)]{Bressan:92} Bressan, A. 1992,
  Mem.\ Soc.\ Astr.\ It., 63, 25

\bibitem[Bressan et al.(2012)]{Bressan:12} Bressan, A., Marigo, P.,
  Girardi, L., Salasnich, B., Dal Cero, C., Rubele, S., \& Nanni,
  A. 2012, \mnras, 427, 127

\bibitem[Claret(2004)]{Claret:04} Claret, A. 2004, \aap, 424, 919

\bibitem[Claret(2007)]{Claret:07} Claret, A. 2007, \aap, 475, 1019

\bibitem[Claret(2012)]{Claret:12} Claret, A. 2012, \aap, 541, 113


\bibitem[Claret \& Gim\'enez(1991)]{Claret:91} Claret, A., Gim\'enez,
  A. 1991, \aap, 244, 319

\bibitem[Claret \& Gim\'enez(2010)]{Claret:10} Claret, A., Gim\'enez,
  A. 2010, \aap, 519, 57


\bibitem[D'Antona et al.(2013)]{DAntona:13} D'Antona, F., Caloi, V.,
  D'Erole, A., Tailo, M., Vesperini, E., Ventura, P., \& Di
  Criscienzo, M. 2013, \mnras, 434, 1138

\bibitem[Deheuvels et al.(2016)]{Deheuvels:16} Deheuvels, S., Brandao,
  I., Silva Aguirre, V.\ et al. 2016, arViv: 1601.01535

\bibitem[Demarque et al.(2004)]{Demarque:04} Demarque, P., Woo, J.-H.,
  Kim, Y.-C., \& Yi, S.\ K. 2004, \apjs, 155, 667

\bibitem[Elgueta et al.(2016)]{Elgueta:16} Elgueta, S.\ S., Graczyk,
  D., Gieren, W.\ et al.\ 2016, \aj, in press

\bibitem[Fekel et al.(2013)]{Fekel:13} Fekel, F.\ C., Henry, G.\ W.,
  \& Sowell, J.\ R. 2013, \aj, 146, 146

\bibitem[Ferguson et al.(2005)]{Ferguson:05} Ferguson, J.\ W.,
  Alexander, D.\ R., Allard, F., Barman, T., Bodnarik, J.\ G.,
  Hauschildt, P.\ H., Heffner-Wong, A., \& Tamanai, A. 2005, \apj, 623,
  585

\bibitem[Gieren et al.(2015)]{Gieren:15} Gieren, W., Pilecki, B.,
  Pietrzy{\'n}ski, G., et al.\ 2015, \apj, 815, 28

\bibitem[Graczyk et al.(2012)]{Graczyk:12} Graczyk, D., Pietrzy\'nski,
  G., Thompson, I.\ B.\ et al. 2012, \apj, 750, 144

\bibitem[Graczyk et al.(2014)]{Graczyk:14} Graczyk, D., Pietrzy\'nski,
  G., Thompson, I.\ B.\ et al. 2014, \apj, 780, 59

\bibitem[Grevesse \& Sauval(1998)]{Grevesse:98} Grevesse, N., \&
  Sauval, A.\ J. 1998, Space Sci.\ Rev., 85, 161

\bibitem[He{\l}miniak et al.(2015)]{Helminiak:15} He{\l}miniak,
  K.\ G., Graczyk, D., Konacki, M., et al.\ 2015, \mnras, 448, 1945

\bibitem[Iglesias \& Rogers(1996)]{Iglesias:96} Iglesias, C.\ A., \&
  Rogers, F.\ J. 1996, \apj, 464, 943

\bibitem[Sandberg Lacy \& Fekel(2011)]{Lacy:11} Sandberg Lacy, C.\ H.,
  \& Fekel, F.\ C. 2011, \aj, 142, 185

\bibitem[Lastennet \& Valls-Gabaud(2002)]{Lastennet:02} Lastennet, E.,
  \& Valls-Gabaud, D. 2002, \aap, 396, 551

\bibitem[Meng \& Zhang(2014)]{Meng:14} Meng, Y., \& Zhang,
  Q.\ S. 2014, \apj, 787, 127

\bibitem[Mowlavi et al.(2012)]{Mowlavi:12} Mowlavi, N., Eggenberger,
  P., Meynet, G., Ekstr\"om, S., Georgy, C., Maeder, A., Charbonnel,
  C., \& Eyer, L. 2012, \aap, 541, 41

\bibitem[Nieuwenhuijzen \& de Jagger(1990)]{Nieuwenhuijzen:90}
  Nieuwenhuijzen, H., \& de Jagger, C. 1990, \aap, 231, 134

\bibitem[Pavlovski et al.(2014)]{Pavlovski:14} Pavlovski, K.,
  Southworth, J., Kolbas, V., \& Smalley, B. 2014, \mnras, 438, 590

\bibitem[Paxton et al.(2011)]{Paxton:11} Paxton, B., Bildsten, L.,
  Dotter, A.\ et al. 2011, \apjs, 192, 3

\bibitem[Pietrinferni et al.(2004)]{Pietrinferni:04} Pietrinferni, A.,
  Cassisi, S., Salaris, M., \& Castelli, F. 2004 \apj, 612, 168

\bibitem[Pietrzy\'nski et al.(2013)]{Pietrzynski:13} Pietrzy\'nski,
  G., Graczyk, D., Gieren, W.\ et al. 2013, \nat, 495, 76

\bibitem[Pietrzy\'nski et al.(2010)]{Pietrzynski:10} Pietrzy\'nski,
  Thompson, I.\ B., Gieren, W.\ et al. 2010, \nat, 468, 542

\bibitem[Pilecki et al.(2013)]{Pilecki:13} Pilecki, B., Graczyk, D.,
  Pietrzy{\'n}ski, G., et al.\ 2013, \mnras, 436, 953

\bibitem[Pilecki et al.(2015)]{Pilecki:15} Pilecki, B., Graczyk, D.,
  Gieren, W., et al.\ 2015, \apj, 806, 29

\bibitem[Pols et al.(1995)]{Pols:95} Pols, O.\ R., Tout, C.\ A.,
  Eggleton P.\ P., Han, Z. 1995, \mnras, 274, 964

\bibitem[Reimers(1977)]{Reimers:77} Reimers, D. 1977, \aap, 61, 217

\bibitem[Ribas(1999)]{Ribas:99} Ribas, I. 1999, Ph.D. Thesis,
  Universitat de Barcelona

\bibitem[Ribas et al.(2000)]{Ribas:00} Ribas, I., Jordi, C., \&
  Gim\'enez, A. 2000, \mnras, 318, 55

\bibitem[Roxburgh(1965)]{Roxburgh:65} Roxburgh, I.\ W. 1965, \mnras,
  130, 223

\bibitem[Roxburgh(1978)]{Roxburgh:78} Roxburgh, I.\ W. 1978, \aap, 65,
  281

\bibitem[Roxburgh(1989)]{Roxburgh:89} Roxburgh, I.\ W. 1989, \aap,
  211, 361

\bibitem[Saslaw \& Schwarzschild(1965)]{Saslaw:65} Saslaw, W.\ C., \&
  Schwarzschild, M. 1965, \apj, 142, 1468

\bibitem[Schaller et al.(1992)]{Schaller:92} Schaller, G., Schaerer,
  D., Meynet, G., \& Maeder, A. 1992, \aaps, 96, 269

\bibitem[Stancliffe et al.(2015)]{Stancliffe:15} Stancliffe, R.\ J.,
  Fossati, L., Passy, J.-C., \& Schneider, F.\ R.\ N. 2015, \aap, 575,
  117

\bibitem[Stothers \& Chin(1991)]{Stothers:91} Stothers, R.\ B., \&
  Chin, C.-W. 1991, \apj, 381, L67

\bibitem[Stothers \& Chin(1992)]{Stothers:92} Stothers, R.\ B., \&
  Chin, C.-W. 1992, \apj, 390, 136

\bibitem[Tkachenko et al.(2014)]{Tkachenko:14} Tkachenko, A.,
  Degroote, P., Aerts, C.\ et al.\ 2014, \mnras, 438, 3093


\bibitem[Torres et al.(2010)]{Torres:10} Torres, G., Andersen, J., \&
  Gim\'enez, A. 2010, \aapr, 18, 67

\bibitem[Torres et al.(2015)]{Torres:15} Torres, G., Claret, A.,
  Pavlovski, K., \& Dotter, A. 2015, \apj, 807, 26

\bibitem[Torres et al.(2014)]{Torres:14} Torres, G., Vaz, L.\ P.\ R.,
  Lacy, C.\ H.\ S., \& Claret, A. 2014, \aj, 147, 36

\bibitem[Valle et al.(2016)]{Valle:16} Valle, G., Dell'Omodarme, M.,
  Prada Moroni, P.\ G., \& Degl'Innocenti, S. 2016, \aap, 587, 16

\bibitem[VandenBerg et al.(2006)]{VandenBerg:06} VandenBerg, D.\ A.,
  Bergbusch, P.\ A., \& Dowler, P. 2006, \apjs, 162, 375

\end{thebibliography}
\end{document}